\RequirePackage{lineno}
\documentclass[aps,prd,twocolumn,superscriptaddress]{revtex4-1}

\usepackage{amssymb}
\usepackage{amsmath}

\usepackage{graphicx}
\usepackage{dcolumn}
\usepackage{bm}
\usepackage{pstricks,pst-node,pst-text,pst-3d}

\usepackage{psfrag}
\usepackage{xspace}

\usepackage{natbib}
\usepackage{url}

\providecommand{\Ks}{K_\mathrm{S}^0}
\providecommand{\KS}{K_\mathrm{S}^0}
\providecommand{\pKs}{pK_\mathrm{S}^0}

\usepackage{ifpdf}
\ifpdf
\DeclareGraphicsExtensions{.pdf, .jpg, .tif}
\usepackage[%
  pdftitle={Pentaquark Theta search at HERMES},%
  pdfauthor={The HERMES Collaboration},%
  pdfsubject={HERMES exotics},%
  pdfstartview=FitH,%
  bookmarks=true,%
  bookmarksopen=true,%
  breaklinks=true,%
  colorlinks=true,%
  linkcolor=blue,anchorcolor=blue,%
  citecolor=blue,filecolor=blue,%
  menucolor=blue,pagecolor=blue,%
  urlcolor=blue]{hyperref}
\else
\DeclareGraphicsExtensions{.eps, .jpg}
\usepackage[%
  breaklinks=true,%
  colorlinks=true,%
  linkcolor=blue,anchorcolor=blue,%
  citecolor=blue,filecolor=blue,%
  menucolor=blue,pagecolor=blue,%
  urlcolor=blue]{hyperref}
\fi

\begin{document}

\title{
  Pentaquark $\Theta^+$ search at HERMES
}


\def\groupargonne{\affiliation{Physics Division, Argonne National Laboratory, Argonne, Illinois 60439-4843, USA}}
\def\groupbari{\affiliation{Istituto Nazionale di Fisica Nucleare, Sezione di Bari, 70124 Bari, Italy}}
\def\groupbeijing{\affiliation{School of Physics, Peking University, Beijing 100871, China}}
\def\groupbilbao{\affiliation{Department of Theoretical Physics, University of the Basque Country UPV/EHU, 48080 Bilbao, Spain and IKERBASQUE, Basque Foundation for Science, 48013 Bilbao, Spain}}
\def\groupcolorado{\affiliation{Nuclear Physics Laboratory, University of Colorado, Boulder, Colorado 80309-0390, USA}}
\def\groupdesy{\affiliation{DESY, 22603 Hamburg, Germany}}
\def\groupzeuthen{\affiliation{DESY, 15738 Zeuthen, Germany}}
\def\groupdubna{\affiliation{Joint Institute for Nuclear Research, 141980 Dubna, Russia}}
\def\grouperlangen{\affiliation{Physikalisches Institut, Universit\"at Erlangen-N\"urnberg, 91058 Erlangen, Germany}}
\def\groupferrara{\affiliation{Istituto Nazionale di Fisica Nucleare, Sezione di Ferrara and Dipartimento di Fisica e Scienze della Terra, Universit\`a di Ferrara, 44122 Ferrara, Italy}}
\def\groupfrascati{\affiliation{Istituto Nazionale di Fisica Nucleare, Laboratori Nazionali di Frascati, 00044 Frascati, Italy}}
\def\groupgent{\affiliation{Department of Physics and Astronomy, Ghent University, 9000 Gent, Belgium}}
\def\groupgiessen{\affiliation{II. Physikalisches Institut, Justus-Liebig Universit\"at Gie{\ss}en, 35392 Gie{\ss}en, Germany}}
\def\groupglasgow{\affiliation{SUPA, School of Physics and Astronomy, University of Glasgow, Glasgow G12 8QQ, United Kingdom}}
\def\groupillinois{\affiliation{Department of Physics, University of Illinois, Urbana, Illinois 61801-3080, USA}}
\def\groupmichigan{\affiliation{Randall Laboratory of Physics, University of Michigan, Ann Arbor, Michigan 48109-1040, USA }}
\def\groupmoscow{\affiliation{Lebedev Physical Institute, 117924 Moscow, Russia}}
\def\groupnikhef{\affiliation{National Institute for Subatomic Physics (Nikhef), 1009 DB Amsterdam, The Netherlands}}
\def\groupstpetersburg{\affiliation{B.P. Konstantinov Petersburg Nuclear Physics Institute, Gatchina, 188300 Leningrad Region, Russia}}
\def\groupprotvino{\affiliation{Institute for High Energy Physics, Protvino, 142281 Moscow Region, Russia}}
\def\grouprome{\affiliation{Istituto Nazionale di Fisica Nucleare, Sezione di Roma, Gruppo Collegato Sanit\`a and Istituto Superiore di Sanit\`a, 00161 Roma, Italy}}
\def\grouptriumf{\affiliation{TRIUMF, Vancouver, British Columbia V6T 2A3, Canada}}
\def\grouptokyo{\affiliation{Department of Physics, Tokyo Institute of Technology, Tokyo 152, Japan}}
\def\groupamsterdam{\affiliation{Department of Physics and Astronomy, VU University, 1081 HV Amsterdam, The Netherlands}}
\def\groupwarsaw{\affiliation{National Centre for Nuclear Research, 00-689 Warsaw, Poland}}
\def\groupyerevan{\affiliation{Yerevan Physics Institute, 375036 Yerevan, Armenia}}
\def\groupnone{\noaffiliation}


\groupargonne
\groupbari
\groupbeijing
\groupbilbao
\groupcolorado
\groupdesy
\groupzeuthen
\groupdubna
\grouperlangen
\groupferrara
\groupfrascati
\groupgent
\groupgiessen
\groupglasgow
\groupillinois
\groupmichigan
\groupmoscow
\groupnikhef
\groupstpetersburg
\groupprotvino
\grouprome
\grouptriumf
\grouptokyo
\groupamsterdam
\groupwarsaw
\groupyerevan

\author{N.~Akopov}  \groupyerevan
\author{Z.~Akopov}  \groupdesy
\author{W.~Augustyniak}  \groupwarsaw
\author{R.~Avakian}  \groupyerevan
\author{A.~Avetissian}  \groupyerevan
\author{E.~Avetisyan}  \groupdesy
\author{S.~Belostotski} \groupstpetersburg
\author{H.P.~Blok}  \groupnikhef \groupamsterdam
\author{A.~Borissov}  \groupdesy
\author{J.~Bowles}  \groupglasgow
% ...

\collaboration{The HERMES Collaboration} \noaffiliation

\date{\today}


\begin{abstract}
The earlier search at HERMES for narrow baryon states excited in quasi-real photoproduction, decaying through the channel $p\KS\rightarrow p\pi^+\pi^-$, has been extended with improved decay-particle reconstruction, more advanced particle identification, and increased event samples. The structure observed earlier at an invariant mass of $1528$\:MeV shifts to $1522$\:MeV and the statistical significance drops to about 2$\sigma$ for data taken with a deuterium target. The number of events above background is $68_{-31}^{+98}\text{(stat)}\pm13\text{(sys)}$.
No such structure is observed in the hydrogen data set.

\begin{description}
\item[PACS numbers]
12.39.Mk, 13.60.Rj, 14.20.Jn
\item[Keywords]
Glueball and nonstandard multi-quark, Pentaquark,
 Baryon production, Baryons
\end{description}
\end{abstract}

\maketitle

\section{\label{sec:level1} Introduction}

Exotic hadrons consisting of five quarks were proposed on the basis of quark and bag models~\cite{gellman64,Zweig64,Jaff76} in the early days of QCD. Predictions based on the Skyrme model~\cite{Man84,Che85,Pra87,Pra03} generated renewed interest in the possible existence of such manifestly exotic baryon states, and chiral-soliton calculations suggested a narrow resonance at $\sim1530$\:MeV~\cite{Dia97}, named $\Theta^+$. Possible experimental evidence for this state came from the observation of a narrow peak at $1.54\pm0.01$\:GeV in both the $K^-$ and $K^+$ missing-mass spectrum for the $\gamma n \to K^+K^-n$ reaction on $^{12}$C~\cite{SPring8}. This observation provoked a series of reports of experimental sightings and theoretical predictions for such states. Increased experimental scrutiny failed to confirm most of these initial reports, and it is now generally accepted that there is no substantial evidence for the existence of the $\Theta^+$ state~\cite{PDG,MA14}.

Motivated by the early reports of its existence, the HERMES Collaboration undertook a search for the $\Theta^+$ in quasi-real photoproduction off a deuterium target. The reaction searched for was inclusive photoproduction of the $\Theta^+$ followed by the decay $\Theta^+ \to \pKs \to p\pi^+\pi^-$. A narrow structure was observed at $1528$\:MeV with a significance of $3.7\,\sigma$~\cite{HERMES04}. Consequently, in spite of the demise of the $\Theta^+$, it remained of interest to improve the sensitivity of the HERMES data to explore the possibility that the observed structure signals a hitherto unobserved baryon resonance.

This paper presents results of a more precise study of the $p\Ks$ mass region near $1528$\:MeV where a narrow structure in the $M(p\Ks)$ distribution was observed in the earlier HERMES search. In addition to an increased number of events analyzed, the new analysis employed a better track reconstruction algorithm and an improved particle identification technique to extract a much cleaner sample of $\KS$'s. In addition, data obtained on a hydrogen target have been analysed.

\section{The experiment}

HERMES was a fixed-internal-target experiment in which the target, a storage cell, was traversed by the circulating beam of the HERA lepton storage ring~\cite{hermestarg}. The target consisted of an open-ended elliptical
storage cell that was aligned coaxially to the lepton beam. The cell
was fed by polarized or unpolarized gas. The polarized target used an
atomic beam source~\cite{abstarg}, which could produce luminosities of
the order of $10^{31}$ to $10^{32}$ cm$^2$/s. Unpolarized data were
obtained using an unpolarized gas feed system, operating with up to three
orders of magnitude higher luminosities. An integrated luminosity of
$\sim 500$\,pb$^{-1}$, corresponding to 28.4 million deep-inelastic
scattering (DIS) events, was collected on a longitudinally polarized
(unpolarized) deuterium target over the years 1998-2000
(2006-2007). With the hydrogen target approximately twice the
luminosity was collected, corresponding to 54.9 million DIS events
accumulated over the years 2002-2005 (2006-2007) on a transversely
polarized (unpolarized) target. Polarized data were summed over the spin orientations.

For an overview of the configuration of the experiment
the reader is referred to the earlier HERMES paper
\cite{HERMES04}. Advances in several aspects of the techniques of the
experiment reported there increase the sensitivity in the
search for new baryon resonances. In the original measurement, particle identification
was accomplished with reconstruction of the event response of
the HERMES ring-imaging Cherenkov detector (RICH)~\cite{RICH}
on a track-by-track basis. This approach was dictated by the limited computing capability
available at that time. However, this technique does not account for complications in particle identification
caused by overlapping Cherenkov rings from two or more tracks in the
same detector half. By its nature, the search reported here focuses on events with at least three tracks. In the analysis presented, the defect is remedied by the implementation of a more advanced method of particle identification, in which the response pattern in the RICH is reconstructed with simultaneous generation of the response
to all the tracks present in an event~\cite{EVENT}. In this way, possible track-to-track cross talk is accounted for and the efficiency and purity of the RICH particle identification is improved.

The $\KS$ spectrum is reconstructed with improved resolution and background rejection.
This results from the use of constraints on the track geometry instead of pion identification with the RICH,
and of data reprocessing with a tracking code involving event-level fitting based on a Kalman-filter algorithm~\cite{Fruhwirth:1987fm}, which corrects the tracking parameters for the effects from magnetic
fields and accounts for all detector materials and known mis-alignments. In this case, imposition of collinearity and track-vertex reconstruction generates spectra of $\KS$ of purity superior to that from the earlier measurements.

\section{Event selection}

In the present analysis, the search for the $\Theta^{+}$ is based on the observation
of events in the decay channel $\Theta^{+}\rightarrow p\KS\rightarrow p\,\pi^{+}\pi^{-}$. Hadron tracks are identified with an efficiency greater than 99\% and a lepton contamination of $<$1\%~\cite{PID3PID5} through the
combined response of a transition-radiation detector, a scintillator hodoscope preceded by two radiation lengths of lead~(the pre-shower detector),
a lead-glass calorimeter and the RICH. Only tracks that are within the spatial volume of fiducial limits corresponding to the acceptance of the HERMES spectrometer and within the momentum range
$1-15$\:GeV are accepted. Events selected must have at least three tracks:
one track identified as a proton by the RICH with momentum in the region $4-15$\:GeV,
in which the RICH is able to identify protons, and at least two oppositely charged tracks
not identified as protons. In the subsequent event reconstruction
these tracks are assumed to be pions, i.e., for tracks between $1-4$\:GeV
the pion hypothesis is always applied while it is done so in the region
of $4-15$\:GeV when the RICH does not identify the particle as a proton.

\begin{figure}[h]
\centering
\includegraphics[width=0.499\textwidth]{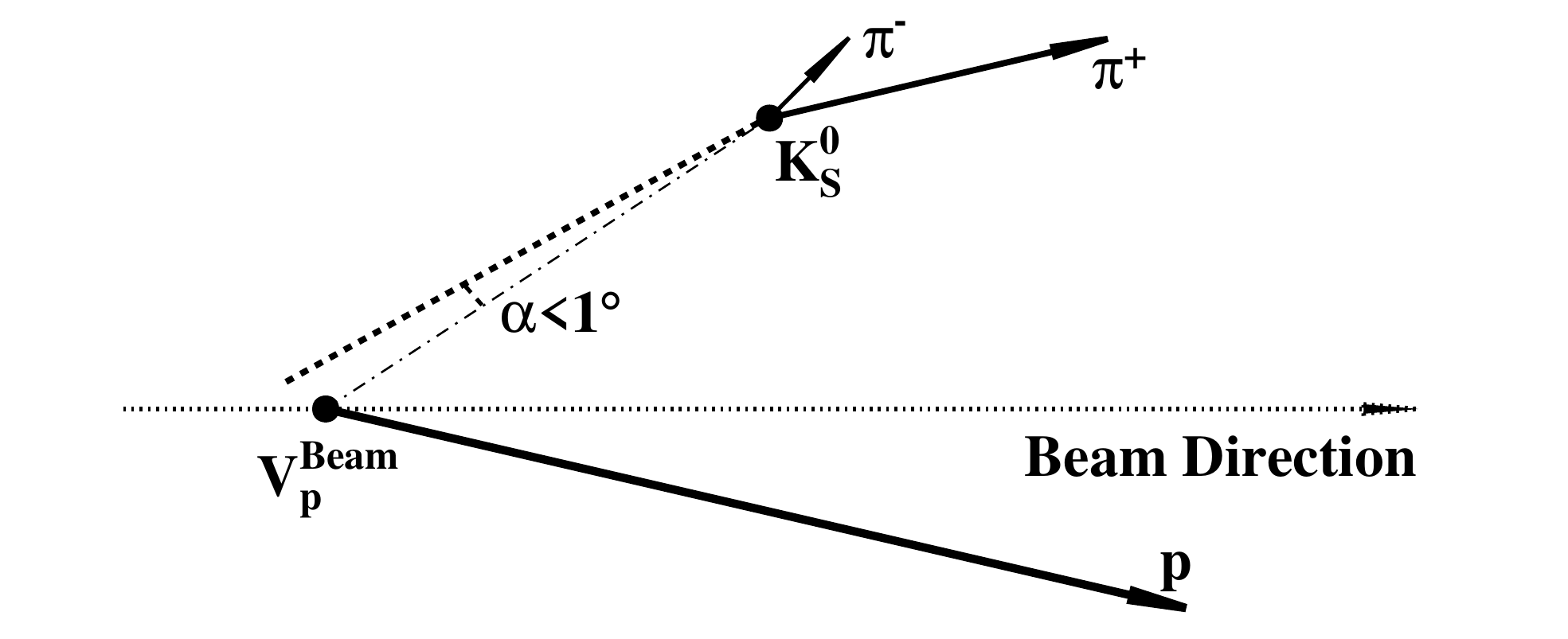}
\caption{Diagram of the kinematic reconstruction of the decay of a $\Theta^{+}$.
The angle $\alpha$ is the difference in the direction of the $\KS$
momentum~(dotted line), as given by the pion momenta, and by the vector
connecting the event origin, $V_p^{\text{Beam}}$, with the decay of the $\KS$~(dash-dotted line).
 \label{FIGpKsMap}}
\end{figure}

The first step of the event reconstruction is the selection of the $\Ks$ through
the invariant-mass spectrum of the two oppositely charged particles, which are
assumed to be pions. The reconstructed trajectory of the $\KS$
candidate is then combined with the proton track to reconstruct the $\Theta^{+}$ candidate. The geometry assumed in the search for the decay of the $\Theta^{+}$ is shown in
Fig.~\ref{FIGpKsMap}.
The momentum of the $\Theta^{+}$ candidate is inferred from the momenta of
the decay pions at their crossing point together with that of the proton.
The $\Theta^{+}$ decay vertex is taken as the intersection of the
proton track with the beam. The distance between the $\KS$ decay point along the beam direction and
the crossing point, $V_p^{\text{Beam}}$, between proton and beam trajectories
must be greater than $4$\:cm. The direction of the momentum of the $\KS$
candidate, as determined by the summed momenta of the decay pions, is
required to agree within one degree with the direction of the vector connecting $V_p^{\text{Beam}}$, assumed to be the production point of the $\Theta^{+}$ candidate, and the point of decay of the $\KS$ ($\alpha<1^{\circ}$ as shown in Fig.~\ref{FIGpKsMap}). The decay vertex of the $\Theta^{+}$ candidate is required to be in the target-cell region, i.e., along the beam direction within ($-20, +20$)\:cm for the long cell used in 1998-2005
and within ($+2, +22$)\:cm for the short cell used in 2006-2007.

\begin{figure}[t]
\centering
\includegraphics[width=0.45\textwidth]{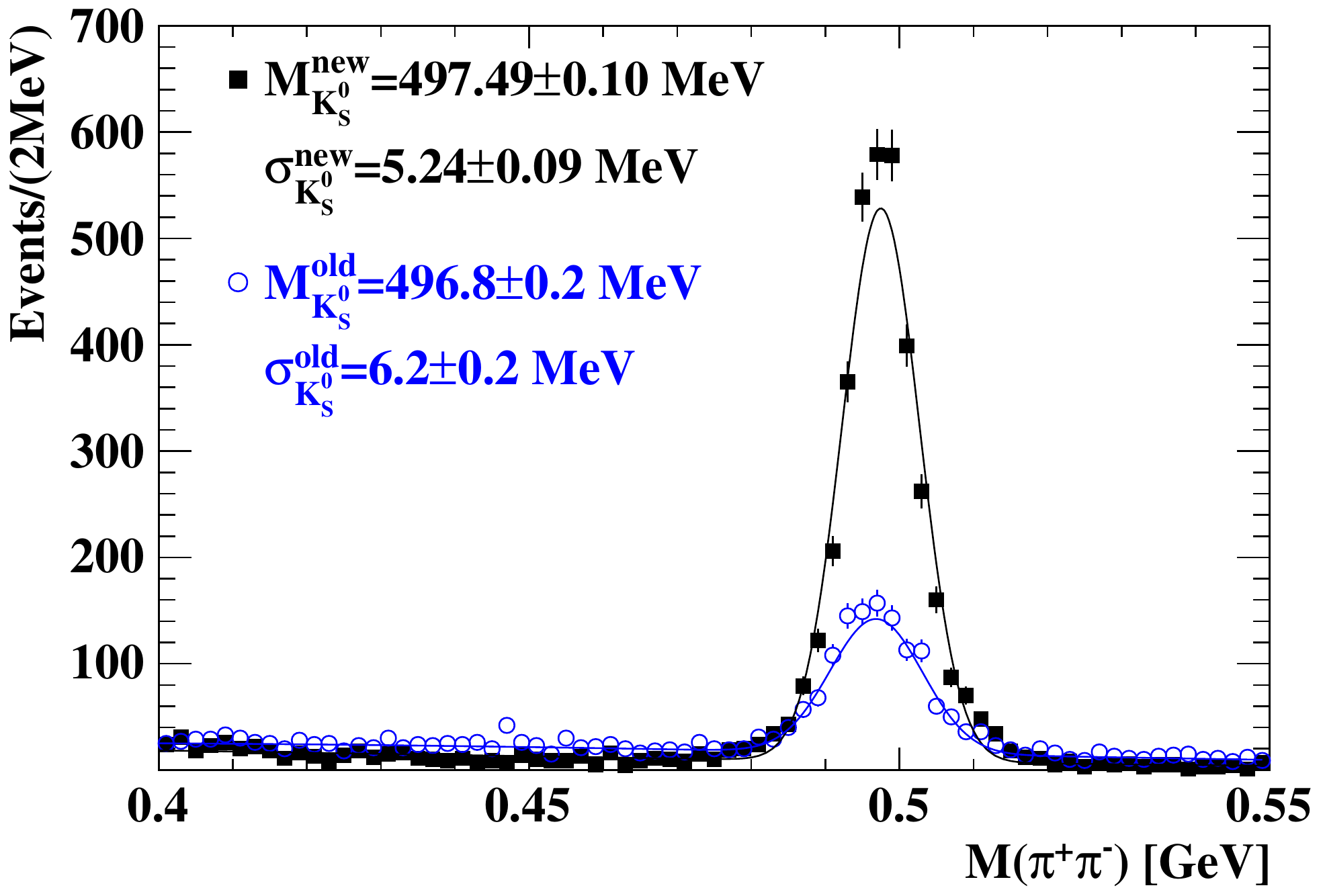}
\caption{
Invariant-mass spectra of two oppositely charged pions showing a clear $\Ks$ signal peak. The filled squares denote this analysis with data from 1998-2000 and 2006-2007 while the open circles are the previously published analysis of the 1998-2000 data. For comparison, the standard deviations and mean values of a single Gaussian function fit to the data together with a third-order Chebychev function for the background are given. The new analysis has a much improved mass resolution and signal-to-noise ratio compared to that of the previous HERMES analysis. \label{FIGKs}}
\end{figure}

The invariant-mass distribution, $M(\pi^+\pi^-)$, of the pion pairs obtained after applying all
selection criteria is shown in Fig.~\ref{FIGKs}. A Gaussian function for the peak together with a
third-order Chebychev function for the background is fitted to the spectrum.
Compared to HERMES data published in 2004~\cite{HERMES04},
the resolution of the $\KS$ peak has been improved from $6.2\pm0.2$\:MeV to
$5.24\pm0.09$\:MeV.
The peak position value agrees within $\pm0.2$\:MeV with the PDG-value $497.614\pm0.024$\:MeV~\cite{PDG}.
The $\KS$ peak is also much cleaner than that of the data in Ref.~\cite{HERMES04}.
The fit as shown in Fig.~\ref{FIGKs} results in the number of $\KS$ of $3311\pm60$ (within $\pm2\sigma$) with $87\pm11$
background events in the new analysis, compared with $963\pm38$ $\KS$ contaminated by $180\pm15$ background events for the previously published $M(\pi^+\pi^-)$ spectrum.

In order to search for the $\Theta^{+}$, events were selected with a $M(\pi^{+}\pi^{-})$
invariant mass within $\pm2\,\sigma$ about the centroid of the $\KS$ peak.

\section{Results}

\begin{figure}[h]
\centering
\includegraphics[width=0.499\textwidth]{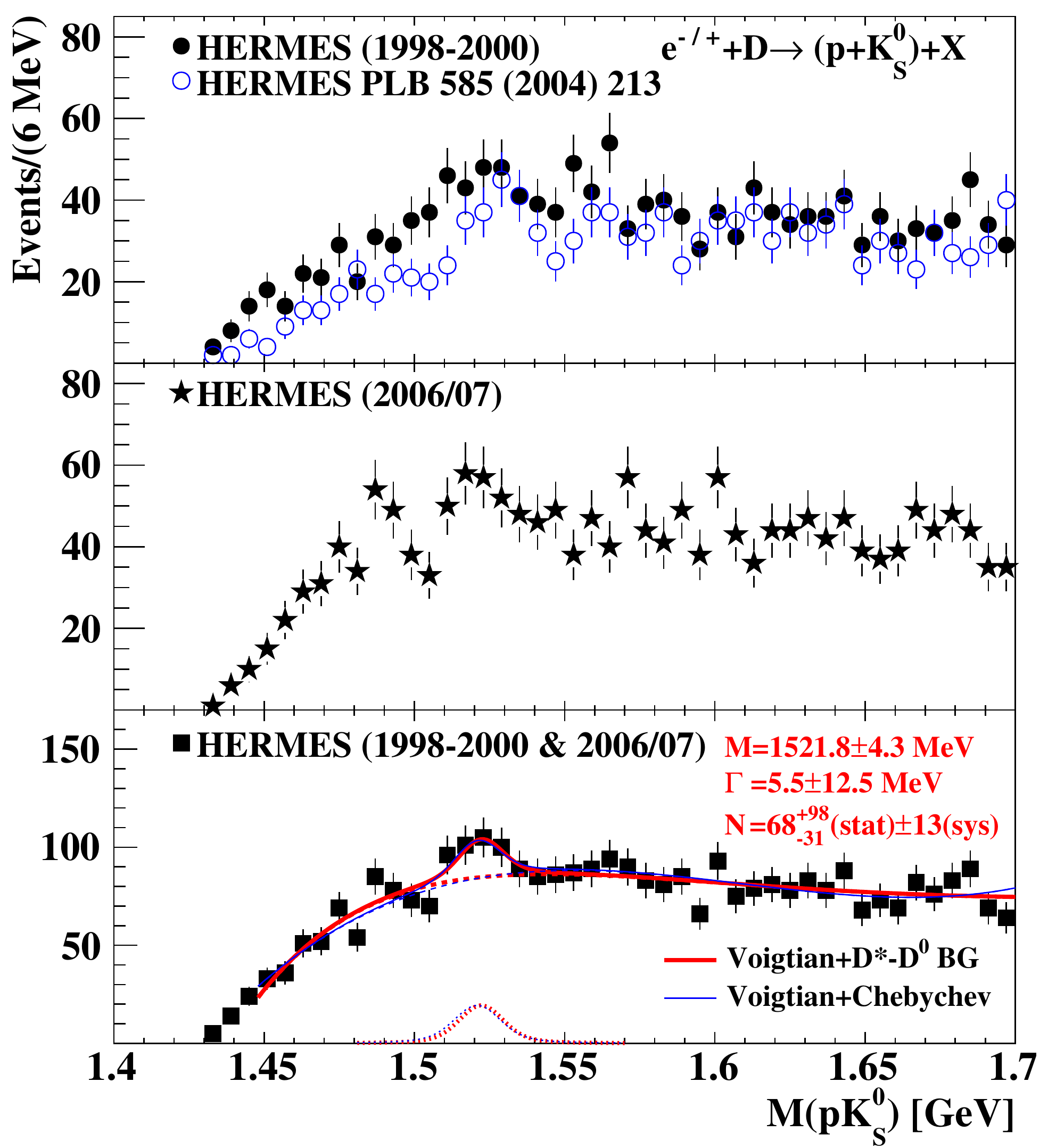}
\caption{
The various $M(pK_S^0)$ spectra for deuterium data taken at the HERMES experiment in the years 1998-2000 (top), 2006-2007 (middle), and for both periods combined (bottom). Also shown in the top panel is the previously published spectrum~\cite{HERMES04} from 1998-2000 of data that has been reanalyzed here. A Voigtian (using a Gaussian with a width fixed to 6\:MeV) together with two different background hypotheses was fitted to the summed spectrum in the bottom panel. The resulting curves are shown separated into signal and background contribution and also combined. The width $\Gamma$ of the Breit--Wigner function, the peak position $M$, and the number of signal events obtained from the fits are given in the panel. \label{FIGThetaRun}
}
\end{figure}

The invariant-mass distributions of the $p\KS$ system, $M(p\KS)$, for data taken with deuterium targets are shown in Fig.~\ref{FIGThetaRun}. It includes the previously published spectrum~(open circles), a spectrum of that data reanalyzed~(filled circles), a spectrum for data taken in the years 2006-2007~(filled stars), and the spectrum resulting from summing the data from both these periods of HERMES running~(filled squares). Only weak suggestions of resonance structure are observed in the newly analyzed spectra.

The presence of significant resonance strength can only be established by a careful analysis. In order to put a limit on the presence of a resonance in the region near $1528$\:MeV reported in the earlier HERMES paper, the summed data were used in a fit of a peak near that energy accompanied by smooth backgrounds. In order to explore the influence of the background shape on the strength of the fitted peak, several different fitting intervals were used. The background shape has been described with the $D^*-D^{0}$ mass-difference function (RooDstD0BG function in RooFit package~\cite{RooFit} of ROOT) and also with a third-order Chebychev shape (RooChebychev function in the RooFit package). The peak function is a Breit--Wigner function convoluted with a Gaussian function (RooVoigtian function in RooFit). The $\sigma$ of the Gaussian function is fixed at $6$\:MeV as determined from a Monte Carlo study of the spectrometer resolution. Fitting the data in different regions yields an average number of signal events $N=68_{-31}^{+98}(\text{stat})\pm13(\text{sys})$. Here, the systematic uncertainty includes the effects of using different background functions and different fit ranges. It also includes the bias determined by repeating many times a Monte Carlo simulation, in which the same statistics as in the real-data spectrum were generated using a fitted shape of the real data. The number of counts under the peak was fitted, and input and output numbers were compared. The average peak position found is $1521.8\pm4.3$\:MeV with a width of the Breit--Wigner function $5.5\pm12.5$\:MeV. A significance of this peak of $1.9\,\sigma$ is obtained from the difference between maximum-likelihood values from un-binned fits~\cite{RooFit} with and without the peak function accompanying a smooth background shape. A value of $2.2\:\sigma$ is obtained when it is estimated from many trials using a Monte Carlo simulation with an event generator giving a smooth shape and each trial fitted with a peak plus background shape, in order to determine the probability to produce a fake peak with a strength equal to or larger than 68, the number of signal events resulting from the fit. Taken together, all these methods show that the significance of a signal for a potential resonant structure at $1521.8$\:MeV is about $2\,\sigma$ for the HERMES deuterium data.

\begin{figure}[t]
\centering
\includegraphics[width=0.48\textwidth]{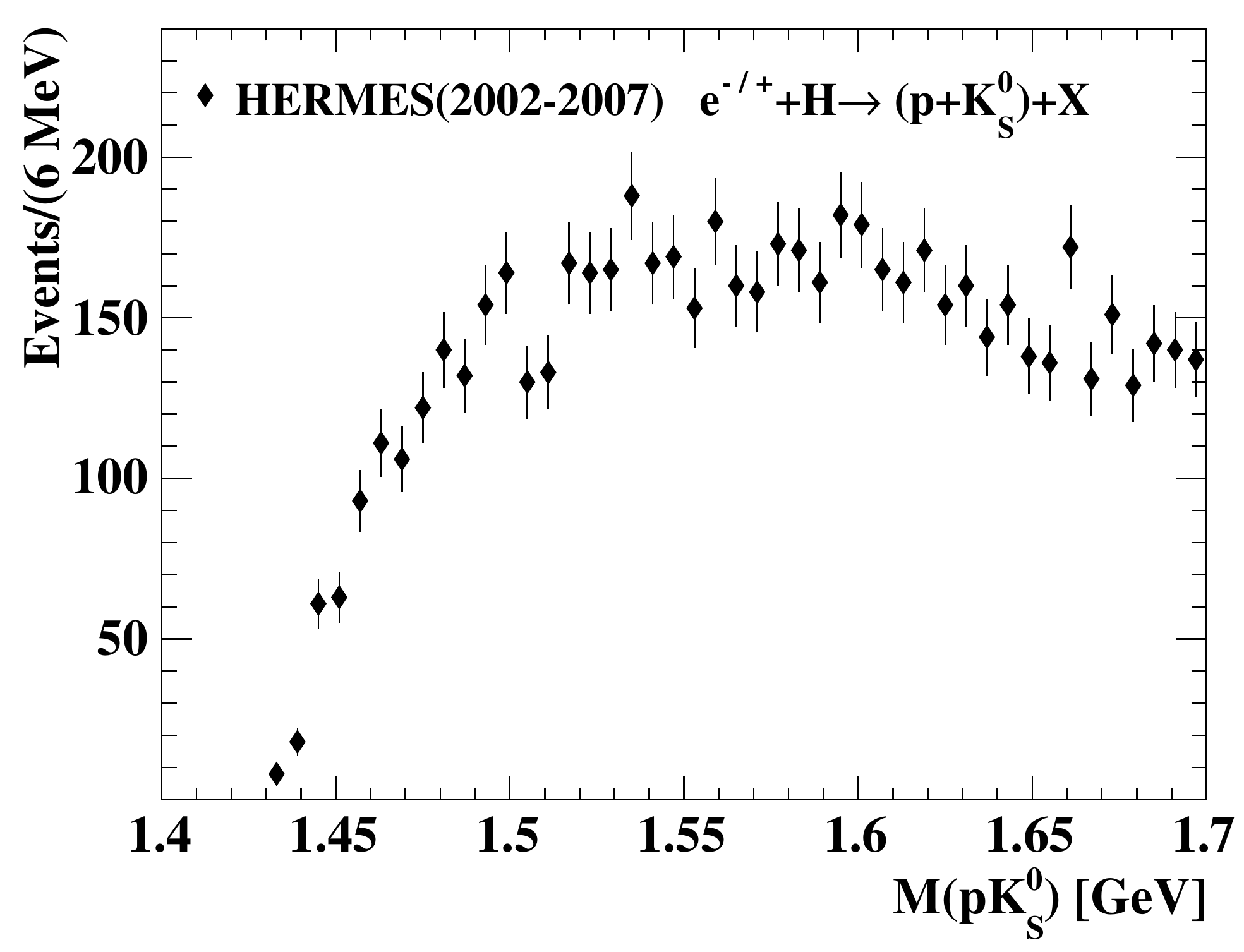}
\caption{ $M(p\KS)$ spectrum from the hydrogen target.
\label{FIG:MpKs_H}}
\end{figure}

For the HERMES hydrogen data there is no evidence for a resonance structure in the $M(p\KS)$ invariant-mass spectrum, as shown in Fig.~\ref{FIG:MpKs_H}.

These results are confirmed by two independent analysis methods~\cite{marianthesis}
based on slightly different event selection criteria. In one case the
events were selected based on a multi-parameter scan in an
optimization of the figure of merit $Z={S}/\sqrt{S+B}$, where $S$ and $B$
refer to the $\Ks$ signal and background yields, respectively, in
the $\Ks$ reconstruction and proton identification. In the other,
the additional requirement of a constrained purity $P=S/(S+B)$ along the
97\% contour in the parameter space was applied. The results for the
$M(p\Ks)$ analysis of all three methods are in statistical agreement
with one another for both targets, hydrogen and deuterium.

\section{Summary}
In summary, the HERMES Collaboration has revisited the earlier reported search~\cite{HERMES04} for a possible $\Theta^+$ excitation in quasi-real photoproduction on a deuterium target with improved tracking and more advanced particle identification. The original data set taken in the years 1998-2000 has been combined with an additional data set taken in the years 2006-2007, resulting in nearly twice as many events as in the original measurement. As a result of the improved tracking and kinematic reconstruction methods, the invariant-mass spectrum of $\KS$ is obtained with significantly less background and better mass resolution. The significance of the potential resonance structure
in the $M(p\KS)$ spectrum of the deuterium data near the $1522$\:MeV region is about $2\,\sigma$, compared to the previously published significance of $3.7\,\sigma$~\cite{HERMES04}. The position of the structure is $6\:$MeV lower in mass than the previously reported $1528\:$MeV, consistent with the accuracies of the old and present analyses.

The observed drop in significance from $3.7\,\sigma$ to about $2\,\sigma$, in spite of twice the number of events for the data from a deuterium target, does not support the presence of a positive $\Theta^{+}$ signal at HERMES kinematics.
For the hydrogen data there is no indication of the existence of an enhancement in the region of interest.
The limited statistics of the HERMES measurement preclude a firm conclusion regarding the existence of five-quark exotic baryons.

\section*{Acknowledgements}

We gratefully acknowledge the DESY management for its support, the staff
at DESY and the collaborating institutions for their significant effort,
as well as our national funding agencies for financial support.

\bibliography{dc75}
\end{document}